\documentclass[a4paper,10pt,twoside]{cpc-hepnp}

\usepackage{multicol}
\usepackage{graphicx}
\usepackage{booktabs}
\usepackage{amssymb,bm,mathrsfs,bbm,amscd}
\usepackage[tbtags]{amsmath}
\usepackage{lastpage}

\begin{document}

\fancyhead[c]{\small Submitted to 'Chinese Physics C'} \fancyfoot[C]{\small \thepage}

\footnotetext[0]{Received 14 March 2009}

\title{Bottomonium states versus recent experimental observations in the QCD-inspired potential model\thanks{Supported by National Natural Science
Foundation of China (11175146, 11047023 and 11265017) and
the Fundamental Research Funds for the Central Universities
(XDJK2012D005)}}

\author{ Tian Wei-Zhao$^{1,2}$
\quad
          CAO Lu$^{1}$
\quad     YANG You-Chang$^{2}$
\quad     CHEN Hong$^{1;1)}$\email{chenh@swu.edu.cn}%
}
 \maketitle

\address{%
$^1$School of Physical Science and Technology, Southwest University, Chongqing 400715, China\\
$^2$Department of Physics, Zunyi Normal College, Zunyi 563002,
China
}

\begin{abstract}
In the QCD-inspired potential model where the quark-antiquark
interaction consists of the usual one-gluon-exchange and the mixture
of long-range scalar and vector linear confining potentials with the
lowest order relativistic correction, we investigate the mass
spectra and electromagnetic processes of a bottomonium system by using the
Gaussian expansion method. It reveals that the vector component
of the mixing confinement is anticonfining and takes around
$18.51\%$ of the confining potential. Combining the new
experimental data released by Belle, BaBar and LHC, we
systematically discuss the energy levels of the bottomonium states
and make the predictions of the electromagnetic decays for further experiments.
\end{abstract}



\begin{multicols}{2}
\section{Introduction}
With the increasing observations of the bottomonium ($b\bar{b}$)
states and bottomonium-like resonances\cite{developments}, it is
worth revisiting the $b\bar{b}$ states to identify the possible
candidates. Very recently, the $\eta_b(2S)$ was firstly observed
in the $\Upsilon(2S)$ radiative decays at a $\sim5 \sigma$ level
\cite{etab2S} at a mass of $9974.6\pm2.3(stat)\pm2.1(syst)$ MeV and
corresponds to the $\Upsilon(2S)$ hyperfine split of
$48.7\pm2.3(stat)\pm2.1(syst)$ MeV. Also as the first observation,
the $h_b(1P)$, $h_b(2P)$\cite{hb} and $\chi_b(3P)$
multiplets\cite{chib3P} have been found recently. Including the
$\Upsilon(10860)$\cite{10860belle2008,10860belle2010} and
$\Upsilon(11020)$\cite{11020a,11020b}, the bottomonium-like states
have aroused lasting interest in theoretical
investigations \cite{screen02,10860bb5S} since their establishment.

The $b\bar{b}$ spectroscopy is considered an excellent laboratory
to examine the quark-antiquark potential within the non-relativistic
framework due to the large $b$-quark mass. The QCD inspired
potential models have played an important role in
investigating the spectra for both charmonium and bottomonium
\cite{QWG169,puzzles}. The long-range confinement is one of the
essential ingredients in most quark potential models. However, the
nature of the confining mechanism is still far away from clear. In the
original Cornell model \cite{cornell01,cornell02}, it was assumed to be
a Lorentz scalar, which gives a vanishing long-range magnetic
contribution and agrees with the flux tube picture of quark
confinement \cite{flux}. Another possibility is that the confinement
may be a more complicated mixture of scalars and time-like vectors,
while the vector potential is anticonfining
\cite{DEbert02,DEbert03}. It has been indicated that the calculated
results of electroweak decay rates for heavy mesons with a pure
scalar confinement are in worse agreement with data than adopting a
vector potential \cite{prd28relativi,mix01,mix02}.

Besides the potential model, the numerical method is very
important for calculating the decays and spectrum of the bound states
system. As several numerical methods fail in the potentials with the
higher than the second orders in $1/r$, the $\mathcal {O}(v^2/c^2)$
corrections to the quark-antiquark potential usually have to be
treated as mass shifts using the leading-order perturbation theory
\cite{prd72,screen02,DEbert03}. Therefore, both perturbation and
nonperturbative treatments have been taken into account recently in
Ref.~\cite{prd75}, which indicates the most significant effect of different treatments on the wave functions. The nonperturbative
treatment brings each state with its own wave function, while the
perturbative treatment leads to the same angular momentum multiplets
sharing the identical wave function. It is known that the radiative
transitions, leptonic and double-photon decay widths are quite
sensitive to the shape of wave function and its information at the
origin. Exactly, each physical particle owning a different quantum
number should behave with a distinguishing state wave function.

We have performed such a nonperturbative treatment on the potential
model calculations for charmonium system in Ref.~\cite{caolu2012cc}
and compared the different confining assumptions. Fitting with the
well-established charmonium states, the Lorentz vector component
parameter is expected to be $22\%$, which implies about one-fifth
vector exchange in the $c\bar{c}$ interquark confining potential.
The subtleness of the obtained wave functions has been examined via
the electromagnetic transition, leptonic decay and two-photon width.
Our predictions for the charmonium are in reasonable agreement with
experiments. It would be interesting and rewarding to extend the
current framework to the bottomonium states in this paper.

In the next section we briefly introduce the QCD-inspired quark
model and the variational approach adopted in this work. Following
with the numerical results, we discuss the latest experimental observations in Section 3. Finally, a summary
is presented in Section 4.

\section{Potential model and calculational approach}

We begin with the nonrelativistic potential model, where the linear
confinement has been assumed as Lorentz scalar-vector mixture
\cite{DEbert02,DEbert03},
\begin{equation}
V_S= \beta(1-\varepsilon)r, \hspace{0.5cm} V_V=-\frac{4}{3}\frac{\alpha _s}{r}+\varepsilon\beta r,
\end{equation}
where $\varepsilon$ stands for the vector exchange scale. Then the
spin-spin, spin-orbit and tensor interactions can be directly
derived from the standard Breit-Fermi expression to order
$(v^2/c^2)$ with the quark mass $m_b$. Explicitly, the adopted
$b\bar{b}$ potential is\cite{caolu2012cc}
\begin{eqnarray}
V_{b\bar{b}}&=& -\frac{4}{3}\frac{\alpha _s}{r}+ \beta\,r+\frac{32\pi \alpha _s}{9m_b^2} \tilde{\delta}_\sigma(r)\bm{S} _b \cdot \bm{S} _{\bar{b}}\nonumber\\
 & &+\left[\frac{2 \alpha _s}{m_b^2r^3}+\frac{(4\varepsilon-1) \beta}{2m_b^2r}\right]\bm{L} \cdot \bm{S}\nonumber\\
 & &+\left[\frac{ \alpha _s}{3m_b^2r^3}+\frac{\varepsilon \beta}{12m_b^2r}\right] \bm{T},
\end{eqnarray}
where $\bm{L}$ is the orbital momentum and $\bm{S}$ is the spin of
bottomonium. The singularity of contact hyperfine interaction within
the spin-spin term has been smeared by Gaussian as
$\tilde{\delta}_\sigma(r)=\left(\sigma/\sqrt{\pi}\right)^3e^{-\sigma^2r^2}$\cite{prd72}.
The involved operators are diagonal in a
$\mid\bm{J},\bm{L},\bm{S}\rangle$ basis with the matrix elements,
\begin{equation}
\langle \bm{S}_b\cdot\bm{S}_{\bar{b}} \rangle = \frac{1}{2} S(S+1)-\frac{3}{4},
\end{equation}
\begin{equation}
\langle \bm{L}\cdot\bm{S}\rangle = \frac{1}{2}\left[J(J+1)-L(L+1)-S(S+1)\right],
\end{equation}
\begin{eqnarray}
\left\langle \bm{T}  \right\rangle &=& \left\langle \left[\frac{3}{r^2}(\bm{S}_b \cdot \bm{r})(\bm{S}_{\bar{b}} \cdot\bm{r})-(\bm{S}_b \cdot \bm{S}_{\bar{b}} )\right] \right\rangle  \\
&=& -\frac{6 \left(\langle \bm{L}\cdot\bm{S}\rangle\right)^2
 + 3 \langle \bm{L}\cdot\bm{S}\rangle - 2S(S+1)L(L+1)}{6(2L-1)(2L+3)}.\nonumber
\end{eqnarray}
Instead of separating the spin-dependent interactions into leading
order parts, we solved the Schr$\ddot{o}$dinger equation of the
unperturbed Hamiltonian with complete $V_{b\bar{b}}(r)$ including
the spin-independent interactions as well as the spin-dependent
terms,
\begin{equation}\label{schr}
\left[-\frac{\hbar^2}{2\mu_R}\nabla^2+V_{b\bar{b}}(r)-E\right]\psi(\textbf{r})=0,
\end{equation}
where $\mu_R= m_b/2$ is the reduced mass. The interaction
Hamiltonian including the kinetic and potential fully enables us
to maintain the subtleness of the wave function. With the help of a
well-chosen set of Gaussian basis functions, namely the Gaussian
Expansion Method \cite{gem}, the singular behavior of $1/r^{3}$ in
spin-dependent terms at short distance can be refined variationally.
The wave functions $\psi_{lm}(\mathbf{r})$ are expanded in terms of
a set of Gaussian basis functions as
\begin{equation}\label{wf1}
\psi_{lm}(\textbf{r})=\sum^{n_{max}}_{n=1}C_{nl}\left(\frac{2^{2l+\frac{7}{2}}\nu_{n}^{l+\frac{3}{2}}}{\sqrt{\pi}(2l+1)!!}\right)^{\frac{1}{2}}
r^{l}e^{-(r/r_{n})^2}Y_{lm}(\hat{\textbf{r}}),
\end{equation}
\begin{equation}\label{wf2}
\nu_{n}=\frac{1}{r_n^2}, \hspace{0.50cm}r_n=r_1a^{n-1}.
\end{equation}
The dimension of Gaussian basis $n_{max}$ is decided by variational
principle. Considering the stableness of the eigen system, we set
the basis-related parameters as $n_{max} = 9$, $r_1 = 0.1$~fm,
$r_{nmax} = 2.2$~fm. In the Gaussian basis space, solving
Eq.(\ref{schr}) involves the generalized eigenvalue problem. The
analytic formulas of matrix elements with the Gaussian basis can be
found in our previous work\cite{caolu2012cc}.

The five model-involved parameters are determined via fitting with
the reasonably well-established $b\bar{b}$ states by minimizing the
merit function
\begin{equation}
\chi^2= \sum^{N}_{i=1}\left[\frac{M^{exp}_{i} - M^{th}_{i}(\mathbf{a})}{\sigma_i}\right]^2,
\end{equation}
where $N$ denotes the number of fitting data,  the $M_{i}^{th}$
indicates the theoretical value, the $M_i^{exp}$ and $\sigma_{i}$ are
the experimental data and the associated errors.  Given a trial set
of model-dependent parameters $\bm{a}$, a procedure calculating the
$\chi^2(\bm{a})$ is developed to improve the trial solution with
the increments $\delta \mathbf{a}$ and repeated until
$\chi^2(\mathbf{a}+\delta \mathbf{a})$ effectively stops decreasing.
The obtained parameters are  $m_b=4.7935$~GeV, $\alpha_s=0.3897$,
$\beta=0.1684$~GeV$^2$, $\sigma=2.1054$~GeV and
$\varepsilon=-0.1851$. The fitted value of $\varepsilon$ implies
that the vector exchange component is approximately one-fifth, which
is close to one obtained for the charmonium system \cite{caolu2012cc}
and consistent with the result of Ref.~\cite{prd75}.

\section{Numerical results}

\subsection{Mass spectrum}
Given the set of parameters, we predict the masses of the forty two
$b\bar{b}$ states shown in Table \ref{bb}, where we compare the
present theoretical predictions with those from the relativized
extension of the nonrelativistic model \cite{prd32} and a
semirelativistic Hamiltonian under nonperturbative framework
\cite{prd75}. Fig. \ref{mass} illustrates the newly found
conventional $b\bar{b}$ states, i.e. $\eta_b(1S)$, $\eta_b(2S)$,
$h_b(1P)$, $h_b(2P)$, $\Upsilon(1^3D_2)$ and the bottomonium-like
$\Upsilon(10860)$ as well as the mass barycenter of the
$\chi_b(3P)$.
\end{multicols}

\begin{center}
\tabcaption{\label{bb} The experimental and theoretical bottomonium mass
spectrums. The labeled states are used in the determination of
potential parameters. We list the world average masses from
PDG\cite{pdg2012} as well as the latest
obersevations\cite{etab2S,hb}.}
\footnotesize
\begin{tabular*}{170mm}{@{\extracolsep{\fill}}cccccc}
 \toprule
   State                     & Expt.~\cite{pdg2012}     & Ref.~\cite{prd32}  & Ref.~\cite{prd75} &  Our    \\
                             &       [MeV]              &        [GeV]      &      [MeV]       &    [MeV]   \\
\hline
   $\eta_b(1^1 S_0)$         & $9390.9\pm 2.8$          &        9.40       &     9421.02      &  9409.17   \\
   $\Upsilon(1^3S_1)^\ast$   & $9460.3\pm0.26$          &        9.46       &     9460.28      &  9458.66   \\
   $\eta_b(2^1 S_0)$       &$9974.6\pm2.3\pm2.1$\cite{etab2S}&   9.98       &     10003.6      &  9996.42   \\
   $\Upsilon(2^3S_1)^\ast$   & $10023.26\pm0.31$        &       10.00       &     10023.5      &  10012.1   \\
   $\eta_b(3^1 S_0)$         &                          &       10.34       &     10360.4      &  10334.9   \\
   $\Upsilon(3^3S_1)^\ast$   &     $10355.2\pm0.5$      &       10.35       &     10365.6      &  10345.5   \\
   $\eta_b(4^1 S_0)$         &                          &                   &     10631.5      &  10612.8   \\
   $\Upsilon(4^3S_1)$        &    $10579.4\pm1.2$       &       10.63       &     10643.4      &  10623.3   \\
   $\eta_b(5^1 S_0)$         &                          &                   &                  &  10865.3   \\
   $\Upsilon(5^3S_1)$        &                          &       10.88       &                  &  10870.6   \\
   $\chi_{b2}(1^3P_2)^\ast$  &  $9912.21\pm0.26$        &        9.90       &     9910.13      &  9924.98   \\
   $\chi_{b1}(1^3P_1)^\ast$  &   $9892.78\pm0.26$       &        9.88       &     9892.83      &  9906.92   \\
   $\chi_{b0}(1^3P_0)^\ast$  &  $9859.44\pm0.42$        &        9.85       &     9860.43      &  9858.59   \\
   $ h_b (1^1P_1)$       &$9898.3\pm1.1^{+1.0}_{-1.1}$\cite{hb}& 9.88       &     9899.94      &  9911.97   \\
   $\chi_{b2}(2^3P_2)^\ast$  &  $10268.65\pm0.22$       &       10.26       &     10271.1      &  10265.6   \\
   $\chi_{b1}(2^3P_1)^\ast$  &  $10255.46\pm0.22$       &       10.25       &     10257.6      &  10250.1   \\
   $\chi_{b0}(2^3P_0)^\ast$  &  $10232.5\pm0.4$         &       10.23       &     10231.4      &  10209.9   \\
   $ h_b (2^1P_1)$     &$10259.8\pm0.6^{+1.4}_{-1.0}$\cite{hb}& 10.25       &     10263.1      &  10254.1   \\
   $\chi_{b2}(3^3 P_2)$      &                          &                   &                  &  10541.4   \\
   $\chi_{b1}(3^3P_1)$       &                          &                   &                  &  10527.6   \\
   $\chi_{b0}(3^3P_0)$       &                          &                   &                  &  10491.3   \\
   $ h_b (3^1P_1)$           &                          &                   &                  &  10531.2   \\
   $\chi_{b2}(4^3 P_2)$      &                          &                   &                  &  10839.9   \\
   $\chi_{b1}(4^3P_1)$       &                          &                   &                  &  10814.9   \\
   $\chi_{b0}(4^3P_0)$       &                          &                   &                  &  10755.2   \\
   $ h_b (4^1P_1)$           &                          &                   &                  &  10814.9   \\
   $\Upsilon(1^3D_3)$        &                          &       10.16       &     10164.1      &  10155.9   \\
   $\Upsilon(1^3D_2)$        &     $10163.7\pm1.4$      &       10.15       &     10157.0      &  10154.4   \\
   $\Upsilon(1^3D_1)$        &                          &       10.14       &     10148.8      &  10149.6   \\
   $1^1D_2$                  &                          &       10.15       &     10158.3      &  10154.2   \\
   $\Upsilon(2^3D_3)$        &                          &       10.45       &     10457.5      &  10441.5   \\
   $\Upsilon(2^3D_2)$        &                          &       10.45       &     10451.2      &  10439.1   \\
   $\Upsilon(2^3D_1)$        &                          &       10.44       &     10443.7      &  10433.7   \\
   $2^1D_2$                  &                          &       10.45       &     10452.4      &  10439.1   \\
   $1^3F_4$                  &                          &       10.36       &     10359.7      &  10337.3   \\
   $1^3F_3$                  &                          &       10.35       &     10355.6      &  10340.1   \\
   $1^3F_2$                  &                          &       10.35       &     10351.0      &  10340.8   \\
   $1^1F_3$                  &                          &       10.35       &     10355.9      &  10339.1   \\
   $2^3F_4$                  &                          &                   &     10617.3      &  10597.3   \\
   $2^3F_3$                  &                          &                   &     10613.4      &  10598.9   \\
   $2^3F_2$                  &                          &                   &     10609.0      &  10598.6   \\
   $2^1F_3$                  &                          &                   &     10613.7      &  10598.2   \\
   \bottomrule
\end{tabular*}%
\end{center}
\begin{multicols}{2}

From the mass spectrum (Table \ref{bb} and Fig. \ref{mass}), it can
be seen that our predictions are in agreement with the experimental
data reasonably. Regarding the $\Upsilon(4S)$, unfortunately, all
the three listed theoretical predications are about 50 MeV higher
than the observation. Since the $\Upsilon(4S)$ is quite close to the
$B\bar{B}$ threshold, it is reasonably shifted towards lower mass by
the meson loops \cite{vanBeveren:2009,vanBeveren:2010ju}. The
threshold effect would make sense on account of $\Upsilon(4S)$
or $\Upsilon(10580)$. $\Upsilon(10860)$ observed by Belle
Collaboration \cite{10860belle2008,10860belle2010} has the world
average mass $M=10.876\pm0.011$ GeV and the full width
$\Gamma=55\pm28$~MeV \cite{pdg2012}. Its structure has been
investigated as the bottomonium $5^3S_1$ state \cite{10860bb5S} and
$b$-flavored $Y(4260)$ \cite{10860belle2008,10860belle2010}. In our
prediction, the theoretical mass $\Upsilon(5S)$
$M(5^3S_1)=10870.61$~MeV is very close to the average value.
However, the bottomonium interpretation of $\Upsilon(10860)$ will be
challenged by the distinguishing partial widths for dipion
transitions \cite{10860belle2010}. The radial excitation of the
$P$-wave $\chi_b$ system was observed in the radiative transitions
to $\Upsilon(1S)$ and $\Upsilon(2S)$ by the ATLAS Collaboration
\cite{chib3P}. The mass barycenter of a spin-triplet was reported as
$10.530\pm0.005\pm0.009$ GeV. Our theoretical prediction is
$\overline{m}_3\equiv\left[m(\chi_{b0})+3m(\chi_{b1})+5m(\chi_{b2})\right]/9=10.5313$~GeV,
which coincides with the experimental data quite well. The
bottomonium ground state $\eta_b(1S)$ was detected by BABAR
Collaboration \cite{etab1S} and confirmed in other experiments. The
world average mass of $\eta_b(1S)$ is $M=9390.9\pm2.8$ MeV and it
corresponds to the $\Upsilon(1S)$ hyperfine split $\Delta
M_{HF}(1S)=69.3\pm2.8$~MeV\cite{pdg2012}. In the present theoretical
framework, we predict $M_{th}=9409.17$~MeV and $\Delta
M_{HF}(1S)\approx 50$~MeV.

\end{multicols}
\ruleup
\begin{center}
\includegraphics[width=13.3cm]{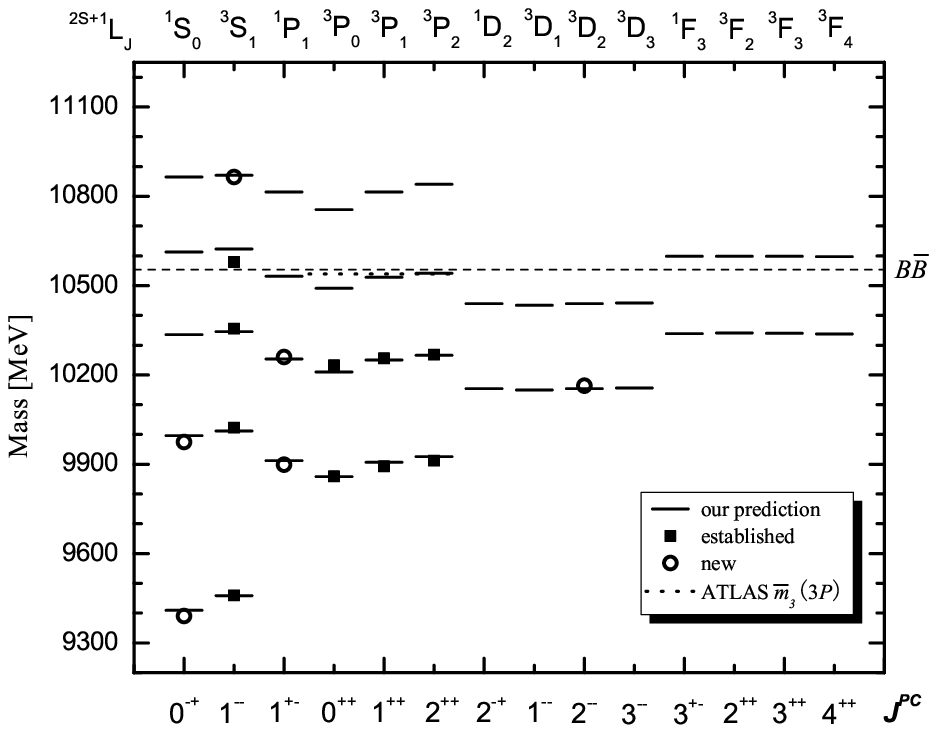}
\figcaption{\label{mass} The established and predicated mass spectrum of the bottomonium states.}
\end{center}
\ruledown
\begin{multicols}{2}

\subsection{Leptonic decays}
The lowest-order expressions of electric decay width with the first-order QCD corrections \cite{prd37} are
\begin{equation}\label{eeS}
\Gamma_{ee} (nS) = \frac{4 \alpha^2 e^2_b}{M^2_{nS}}\mid R_{nS}(0)\mid^2\left(1-\frac{16}{3}\frac{\alpha_s(m_b)}{\pi}\right),
\end{equation}

\begin{equation}\label{eeD}
\Gamma_{ee} (nD) = \frac{25 \alpha^2 e^2_b}{2 M^2_{nS} m^4_{b}}\mid R^{\prime\prime}_{nD}(0)\mid^2\left(1-\frac{16}{3}\frac{\alpha_s(m_b)}{\pi}\right),
\end{equation}
where $e_b$ is the $b$-quark charge in units of $|e|$, $\alpha = 1/137.036$ is the fine-structure constant, $M_{nS}$ and $M_{nD}$ are the $(n_r+1)th$ $S$-wave and D-wave state mass respectively with the radial excitation number $n_r$. Note that here $\alpha_s(m_b)$ within potential are essentially the strong coupling constants but of different mass scales. We adopt $\alpha_s(m_b) = 0.18$ as in Refs. \cite{screen02,DEbert2003twophoton}. $R_{nS}(0)$ is the radial $S$ wave function at the origin, and $R^{\prime\prime}_{nD}(0)$ is the second derivative of the radial $D$-wave function at the origin. They are explicitly analytic in the Gaussian basis space\cite{caolu2012cc}. Table \ref{ee} presents the numerical results.

\end{multicols}
\begin{center}
\tabcaption{\label{ee} The leptonic decay width, in units of keV.}
\footnotesize
\begin{tabular*}{170mm}{@{\extracolsep{\fill}}cccccccc}
 \toprule
          State         &  \multicolumn{2}{c}{Ref.~\cite{screen02}}&Ref.~\cite{bbee161}&Ref.~\cite{bbee162}&\multicolumn{2}{c}{Our}&    Expt.~\cite{pdg2012}\\
          \cline{2-3}\cline{6-7}
                        & $\Gamma^{0}_{ee}$ &   $\Gamma_{ee}$    &  $\Gamma_{ee}$    &  $\Gamma_{ee}$    & $\Gamma^{0}_{ee}$ & $\Gamma_{ee}$ &   \\
\hline
$\Upsilon(1^3S_1)$      & 2.31              & 1.60               & 1.314             &1.320              & 1.738           &  1.207          &1.340$\pm$0.018 \\
$\Upsilon(2^3S_1)$      & 0.92              & 0.64               & 0.576             &0.628              & 0.666           &  0.462          &0.612$\pm$0.011 \\
$\Upsilon(3^3S_1)$      & 0.64              & 0.44               & 0.476             &0.263              & 0.494           &  0.343          & 0.443$\pm$0.08 \\
$\Upsilon(4^3S_1)$      & 0.51              & 0.35               & 0.248             &0.104              & 0.544           &  0.378          &0.272$\pm$0.029 \\
$\Upsilon(5^3S_1)$      & 0.42              & 0.29               & 0.310             &0.04               & 0.272           &  0.189          &                \\
$\Upsilon(1^3D_1)$      &                   &                    &                   &                   & 0.00121         &  0.000838       &                \\
$\Upsilon(2^3D_1)$      &                   &                    &                   &                   & 0.00205         &  0.00143        &                \\
   \bottomrule
\end{tabular*}%
\end{center}
\begin{multicols}{2}
\subsection{Radiative transition}
Because radiative transition is sensitively dependent on the
detailed features of state wave functions, it is of great interest
as a plausible inspection of meson structure. The transition rate
between an initial bottomonium state $i$ of radial quantum number
$n_i$, orbital angular momentum $L_i$, spin $S_i$, and total angular
momentum $J_i$, and a final state $f$ with corresponding labels
$(\hbar=c=1)$ is given in Ref.~\cite{Kwong1988} as
\begin{eqnarray}\label{e1}
&&\Gamma_{E1}\left(n_i^{2S_i+1}L_{i_{J_i}}\rightarrow  n_f^{2S_f+1}L_{f_{J_f}}\right)
=\frac{4}{3}C_{fi}\delta_{S_iS_f}e^2_b\alpha\nonumber\\
& &\times\left|\left\langle \psi_f \left|r\right| \psi_i \right\rangle \right|^2
E^3_{\gamma}\frac{E_f^{(b\bar{b})}}{M_i^{(b\bar{b})}},
\end{eqnarray}
\begin{eqnarray}\label{m1}
&&\Gamma_{M1}\left(n_i^{2S_i+1}L_{i_{J_i}}\rightarrow   n_f^{2S_f+1}L_{f_{J_f}}\right)
=\frac{4}{3}\frac{2J_f+1}{2L_i+1}e^2_b \nonumber \\
&& \times\frac{\alpha}{m_b^2}\delta_{L_iL_f} \delta_{S_i,S_f\pm 1}  \left|\left\langle \psi_f \mid \psi_i \right\rangle \right|^2
 E^3_{\gamma}\frac{E_f^{(b\bar{b})}}{M_i^{(b\bar{b})}}.
\end{eqnarray}

In the above formulas,  $e_b$ is the charge of b-quark in units of $|e|$, and $M_i$, $E_f$ symbolize the eigen mass of initial $b\bar{b}$ and the total energy of final state respectively. The momentum of the final photon equals $E_{\gamma}=(M^2_i-M^2_f)/(2M_i)$ to leading nonrelativistic order \cite{QWG}. The variationally Gaussian expanded wave functions give rise to the analytic formulas for the overlap integral and the transition matrix elements\cite{caolu2012cc}. The angular matrix element $C_{fi}$ is
\begin{equation}
C_{fi}=max(L_i,L_f)(2J_f+1)\left\{
                 \begin{array}{clr}
                  L_f & J_f & S \\
                  J_i & L_i & 1
                 \end{array} \right\}^2.
\end{equation}
The numerical results of $E1$ and $M1$ transitions are presented in
Tables \ref{E1-1}-\ref{E1-2} and \ref{M1}, respectively. Our results
are compatible with the observations for most channels.

\end{multicols}
\begin{center}
\tabcaption{\label{E1-1} E1 radiative transitions }
\footnotesize
\begin{tabular*}{170mm}{@{\extracolsep{\fill}}ccccccccc}
 \toprule
    Initial             & Final  &\multicolumn{3}{c}{$E_{\gamma}$ [MeV]} &\multicolumn{3}{c}{$\Gamma_{E1}$ [keV]}& Expt.\cite{pdg2012}[keV]\\
    \cline{3-5}\cline{6-8}
                        &&Ref.~\cite{Kwong1988}& Ref.~\cite{screen02}& Our    &Ref.~\cite{Kwong1988}& Ref.~\cite{screen02}& Our         &     \\
\hline
 $\Upsilon(2^3S_1)$     &$\chi_{b2}(1^3P_2)$       &110     &110       &110    &2.14    &2.62    &2.64        &2.287$\pm$0.112\\
                        &$\chi_{b1}(1^3P_1)$       &131     &130       &130    &2.18    &2.54    &2.29        &2.207$\pm$0.128\\
                        &$\chi_{b0}(1^3P_0)$       &162     &163       &162    &1.39    &1.67    &1.047       &1.215$\pm$0.128\\
$\eta_b(2^1S_0)$        &$h_b(1^1P_1)$             &        &83        &84     &        &6.10    &2.045       &\\
 $\Upsilon(3^3S_1)$     &$\chi_{b2}(1^3P_2)$       &433     &434       &434    &0.025   &0.25    &0.681       &$<$0.386\\
                        &$\chi_{b1}(1^3P_1)$       &453     &452       &452    &0.017   &0.17    &0.0886      &$<$0.0345\\
                        &$\chi_{b0}(1^3P_0)$       &484     &484       &484    &0.007   &0.007   &0.0495      &0.0610$\pm$0.0224\\
 $\eta_b(3^1S_0)$       &$h_b(1^1P_1)$             &        &418       &414    &        &1.24    &0.997       &\\
 $\Upsilon(3^3S_1)$     &$\chi_{b2}(2^3P_2)$       &86      &          &86     &2.78    &        &3.138       &2.662$\pm$0.325\\
                        &$\chi_{b1}(2^3P_1)$       &99      &          &99     &2.52    &        &2.538       &2.561$\pm$0.244\\
                        &$\chi_{b0}(2^3P_0)$       &124     &          &122    &1.65    &        &1.122       &1.200$\pm$0.122\\
$\eta_b(3^1S_0)$        &$h_b(2^1P_1)$             &        &          &80     &2.14    &        &4.471       &\\
$\chi _{b2}(1^3P_2) $   &$\Upsilon(1^3S_1)$        &443     &442       &455    &37.8    &38.2    &35.739      &\\
$\chi _{b1}(1^3P_1)$    &                          &443     &423       &438    &32.8    &33.6    &33.310      &\\
$\chi _{b0}(1^3P_0)$    &                          &392     &391       &392    &26.1    &26.6    &25.376      &\\
$h_b (1^1P_1)$         &$\eta _b (1^1S_0)$        &        &501       &490    &        &55.8    &39.015      & \\
$\chi _{b2}(2^3 P_2)$ &$\Upsilon(2^3S_1)$        &242     &243       &250    &18.7    &18.8    &17.326      & \\
$\chi _{b1}(2^3P_1)$  &                          &230     &230       &235    &15.9    &15.9    &16.505      & \\
\bottomrule
\end{tabular*}%
\end{center}

\begin{center}
\tabcaption{\label{E1-2}  E1 radiative transitions (continued)}
\footnotesize
\begin{tabular*}{170mm}{@{\extracolsep{\fill}}ccccccccc}
 \toprule
    Initial             & Final  &\multicolumn{3}{c}{$E_{\gamma}$ [MeV]} &\multicolumn{3}{c}{$\Gamma_{E1}$ [keV]}& Expt.~\cite{pdg2012}[keV]\\
       \cline{3-5}\cline{6-8}
                        &&Ref.~\cite{Kwong1988}& Ref.~\cite{screen02}& Our    &Ref.~\cite{Kwong1988}& Ref.~\cite{screen02}& Our         &     \\
\hline
$\chi _{b0}(2^3P_0)$  &                          &205     &207       &196    &11.3    &11.7    &12.092      & \\
  $h_b(2^1P_1)$         &$\eta_b(2^1S_0)$          &        &266       &254    &        &24.7    &17.61       &\\
  $\chi_{b2}(2^3P_2)$   &$\Upsilon(1^3S_1)$        &777     &777       &775    &9.75    &12.0    &13.736 &\\
  $ \chi_{b1}(2^3P_1)$  &                          &765     &764       &761    &9.31    &12.4    &9.618 &\\
   $\chi_{b0} (2^3P_0)$ &                          &742     &743       &724    &8.48    &11.4    &2.354 &\\
  $h_b(2^1P_1)$         &$\eta _b(1^1S_0)$         &        &831       &810    &        &15.9    &14.861 & \\
  $\chi_{b2}(2^3P_2)$   &$\Upsilon(1^3D_3)$        &107     &113       &109    &2.62    &3.33    &2.781 &\\
                        &$\Upsilon(1^3D_2)$        &112     &117       &111    &0.54    &0.66    &0.488 &\\
                        &$\Upsilon(1^3D_1)$        &119     &123       &115    &0.043   &0.05    &0.0342 &\\
  $ \chi_{b1}(2^3P_1)$  &$\Upsilon(1^3D_2)$        &99      &104       &95     &1.86    &2.21    &1.755 &\\
                        &$\Upsilon(1^3D_1)$        &106     &110       &100    &0.76    &0.92    &0.631 &\\
  $\chi_{b0} (2^3P_0)$  &$\Upsilon(1^3D_1)$        &81      &87        &60     &1.36    &1.83    &0.682 &\\
  $h_b(2^1P_1)$         &$h _{b2}(1^1D_2)$         &        &104       &99     &        &7.74    &2.603 & \\
  $\chi _{b2} (3^3 P_2)$&$\Upsilon(3^3S_1)$        &170     &183      &194     &12.1    &15.6    &16.250 & \\
  $\chi _{b1}(3^3P_1)$  &                          &159     &167      &181     &10.1    &12.0    &15.270 & \\
  $\chi _{b0}(3^3P_0)$  &                          &144     &146      &145     &7.46    &7.88    &10.331 & \\
  $h_b(3^1P_1)$         &$\eta_b(3^1S_0)$          &        &196      &194     &        &19.2    &16.286 &\\
  $\chi _{b2} (3^3 P_2)$&$\Upsilon(2^3S_1)$        &491     &504      &516     &3.78    &6.00    &5.923 & \\
  $\chi _{b1}(3^3P_1)$  &                          &481     &489      &503     &3.56    &5.48    &4.192 & \\
  $\chi _{b0}(3^3P_0)$  &                          &466     &468      &468     &3.24    &4.80    &0.762 & \\
  $h_b(3^1P_1)$         &$\eta_b(2^1S_0)$          &        &528      &521     &        &6.89    &5.828 &\\
  $\chi_{b2}(3^3P_2)$   &$\Upsilon(1^3S_1)$        &1012    &1024     &1027    &3.80    &7.09    &7.330 &\\
  $ \chi_{b1}(3^3P_1)$  &                          &1003    &1010     &1015    &3.69    &6.80    &4.238 &\\
  $\chi_{b0} (3^3P_0)$  &                          &989     &990      &982     &3.54    &6.41    &0.217 &\\
  $h_b(3^1P_1)$         &$\eta _b(1^1S_0)$         &        &1078     &1062    &        &8.27    &7.668 & \\
  $\chi_{b2}(3^3P_2)$   &$\Upsilon(2^3D_3)$        &82      &97       &99      &3.01    &5.05    &4.662 &\\
                        &$\Upsilon(2^3D_2)$        &85      &101      &102     &0.61    &1.02    &0.841 &\\
                        &$\Upsilon(2^3D_1)$        &91      &107      &107     &0.05    &0.08    &0.0603&\\
  $ \chi_{b1}(3^3P_1)$  &$\Upsilon(2^3D_2)$        &75      &86       &88      &2.08    &3.10    &3.098 &\\
                        &$\Upsilon(2^3D_1)$        &81      &92       &94      &0.86    &1.26    &1.146 &\\
  $\chi_{b0} (3^3P_0)$  &$\Upsilon(2^3D_1)$        &        &         &57      &        &        &1.348 &\\
  $h_b(3^1P_1)$         &$h _{b2}(2^1D_2)$         &        &         &92      &        &        &4.537 & \\
  $\chi_{b2}(3^3P_2)$   &$\Upsilon(1^3D_3)$        &        &377      &379     &$\approx0$  &$\approx0$  &0.000261 &\\
                        &$\Upsilon(1^3D_2)$        &        &381      &379     &$\approx0$  &$\approx0$  &0.000786 &\\
                        &$\Upsilon(1^3D_1)$        &        &387      &385     &$\approx0$  &$\approx0$  &0.003803&\\
  $ \chi_{b1}(3^3P_1)$  &$\Upsilon(1^3D_2)$        &        &366      &367     &$\approx0$  &$\approx0$  &0.0584&\\
                        &$\Upsilon(1^3D_1)$        &        &372      &371     &$\approx0$  &$\approx0$  &0.00479 &\\
  $\chi_{b0} (3^3P_0)$  &$\Upsilon(1^3D_1)$        &        &336      &336     &$\approx0$  &$\approx0$  &0.957&\\
  $h_b(3^1P_1)$         &$h _{b2}(1^1D_2)$         &        &370      &370     &$\approx0$  &$\approx0$  &0.0384 & \\
 $\Upsilon(1^3D_3)$     & $\chi_{b2}(1^3P_2)$      &245     &240      &228     &24.3    &26.4     &22.060 &\\
  $\Upsilon(1^3D_2)$    & $\chi_{b2}(1^3P_2)$      &240     &236      &227     &5.7     &6.29     &5.438 &\\
                        & $\chi_{b2}(1^3P_1)$      &261     &255      &245     &22.0    &23.8     &18.481 &\\
  $\Upsilon(1^3D_1)$    & $\chi_{b2}(1^3P_2)$      &233     &230      &222     &0.58    &0.65     &0.570 &\\
 $\Upsilon(1^3D_1)$    & $\chi_{b2}(1^3P_1)$      &254     &249      &240     &11.3    &12.3     &9.765 &\\
                        & $\chi_{b2}(1^3P_0)$      &285     &282      &287     &21.4    &23.6     &17.267 &\\
  $h _{b2}(1^1D_2)$     & $h_b(1^1P_1)$            &        &246      &239     &        &42.3     &23.780 & \\
  $\Upsilon(2^3D_3)$    & $\chi_{b2}(1^3P_2)$      &518     &517      &504     &3.94    &4.01     &3.042 &\\
  $\Upsilon(2^3D_2)$    & $\chi_{b2}(1^3P_2)$      &514     &513      &501     &0.97    &0.98     &0.551 &\\
                        & $\chi_{b2}(1^3P_1)$      &534     &531      &519     &3.25    &3.26     &3.285 &\\
  $\Upsilon(2^3D_1)$    & $\chi_{b2}(1^3P_2)$      &509     &507      &496     &0.10    &0.11     &0.0353 &\\
                        & $\chi_{b2}(1^3P_1)$      &529     &525      &513     &1.75    &1.76     &1.266 &\\
                        & $\chi_{b2}(1^3P_0)$      &559     &557      &559     &2.76    &2.79     &5.208 &\\
  $h _{b2}(2^1D_2)$     & $h_b(1^1P_1)$            &        &522      &514     &        &6.19     &3.853 &\\
\bottomrule
\end{tabular*}%
\end{center}
\newpage


\begin{center}
\tabcaption{\label{M1} M1 radiative partial widths }
\footnotesize
\begin{tabular*}{170mm}{@{\extracolsep{\fill}}ccccc}
 \toprule
 Initial   & Final             & $E_{\gamma}$ [MeV]& $\Gamma_{M1}$ [keV]  &Expt.~\cite{pdg2012} [keV]\\
\hline
$\Upsilon(1^3S_1)$             &$\eta _b (1^1S_0)$         & 49  &5.605 &               \\
$\Upsilon(2^3S_1)$             &$\eta_b(2^1S_0)$           & 16  &0.180 &               \\
                               &$\eta _b(1^1S_0)$          & 585 &19.933& $12.47\pm4.80$ \\
$\eta_b(2^1S_0)$               &$\Upsilon(1^3S_1)$         & 523 &46.385&                \\
$\Upsilon(3^3S_1)$             &$\eta_b(3^1S_0)$           & 11  &0.0556&                \\
                               &$\eta_b(2^1S_0)$           & 343 &2.515 & $< 12.60$      \\
                               &$\eta _b (1^1S_0)$         & 894 &19.029& $10.36\pm1.42$  \\
$\eta_b(3^1S_0)$               &$\Upsilon(2^3S_1)$         & 318 &6.806 &                \\
                               &$\Upsilon(1^3S_1)$         & 839 &45.005&                 \\
$h_b(2^1P_1)$                  &$\chi_{b2}(1^3P_2)$        & 324 &3.806 &                \\
                               &$\chi_{b1}(1^3P_1)$        & 341 &0.449 &                 \\
                               &$\chi_{b0}(1^3P_0)$        & 388 &14.062&                \\
 $\chi_{b2}(2^3P_2)$           &$h_b(1^1P_1)$              & 347 &2.566 &               \\
 $\chi_{b1}(2^3P_1)$           &$h_b(1^1P_1)$              & 333 &0.431 &              \\
 $\chi_{b0}(2^3P_0)$           &$h_b(1^1P_1)$              & 294 &25.509&         \\
 $h_b(3^1P_1)$                 &$\chi_{b2}(2^3P_2)$        & 262 &2.653 &             \\
                               &$\chi_{b1}(2^3P_1)$        & 277 &0.277 &          \\
                               &$\chi_{b0}(2^3P_0)$        & 316 &8.604 &              \\
                               &$\chi_{b2}(1^3P_2)$        & 589 &5.286 &             \\
                               &$\chi_{b1}(1^3P_1)$        & 606 &0.600 &          \\
                               &$\chi_{b0}(1^3P_0)$        & 651 &19.314&          \\
 $\chi_{b2}(3^3P_2)$           &$h_b(2^1P_1)$              & 283 &1.745 &          \\
                               &$h_b(1^1P_1)$              & 611 &3.578 &           \\
 $\chi_{b1}(3^3P_1)$           &$h_b(2^1P_1)$              & 270 &0.271 &           \\
                               &$h_b(1^1P_1)$              & 598 &0.572 &          \\
 $\chi_{b0}(3^3P_0)$           &$h_b(2^1P_1)$              & 235 &17.450&            \\
                               &$h_b(1^1P_1)$              & 563 &35.015&        \\
\bottomrule
\end{tabular*}%
\end{center}

\begin{multicols}{2}
\section{Summary\label{4}}
The bottomonium mass spectrum, electromagnetic transitions and
leptonic widths are investigated by adopting the quark-antiquark
potential consisting of the one-gluon-exchange and the Lorentz
scalar-vector mixing linear confining potentials. We preform a
nonperturbative calculation to the full Hamiltonian including
spin-independent and -dependent potentials. Fitting with the mass of
the ten well-established $b\bar{b}$ states, the vector scale
parameter $\varepsilon$ is determined and implies the $18.51\%$
vector component of confining interaction. Combining our previous
work on charmonium systems\cite{caolu2012cc}, we found that the
scalar-vector mixing linear confinement of heavy quarkonium seems to
be important and nearly about one-fifth; explicitly, it is slightly
lower than $4\%$ for the bottomonium mesons.

In our calculation, the mass spectrum accords reasonably well with
the newly observed family members of bottomonium, i.e. $\eta_b(1S)$,
$\eta_b(2S)$, $h_b(1P)$, $h_b(2P)$, $\Upsilon(1^3D_2)$ and the
$\chi_b(3P)$ structure. The mass and leptonic decay width of
$5^3S_1$ is very close to the data of the bottomonium-like
$\Upsilon(10860)$ released by Belle in our calculation and Ref
\cite{bbee161}, respectively. However, the bottomonium
interpretation of $\Upsilon(10860)$ will be challenged by the
distinguishing partial widths for dipion transitions
\cite{10860belle2010}. Very recently, the production rates have been
calculated by Ahmed and Wang \cite{Ali:2011qi} for the processes
$pp(\bar{p})\rightarrow
Y_b(10890)(\rightarrow\Upsilon(1S,2S,3S)\pi^+\pi^-\rightarrow\mu^+\mu^-\pi^+\pi^-)$,
by taking the $Y_b(10890)$ as a tetraquark state. Further experiment
at the LHC and the Tevatron will help us to understand its
structure. The corresponding $E1$, $M1$ radiative transitions and
leptonic decay widths have been theoretically predicted for the
further experimental search and enrichment of the bottomonium
family.

\vspace{-1mm}
\centerline{\rule{80mm}{0.1pt}}

\end{multicols}


\clearpage

\end{document}